\documentclass[seceq,supplement]{ptptex}

\newcommand{\ave}[1]{\left\langle #1 \right\rangle}
\newcommand{\qln}{ \ln_{q} }
\newcommand{\qexp}{ \exp_{q} }

\title{
The Boltzmann temperature and Lagrange multiplier in
nonextensive
thermostatistics%
}

\author{
Tatsuaki \textsc{Wada}$^1$ and Antonio M. \textsc{Scarfone}$^2$%
}

\inst{
$^1$ Department of Electrical and Electronic Engineering,
Ibaraki University, Ibaraki, 316-8511, Japan
\\
$^2$ Dipartimento di Fisica, Politecnico di Torino, Corso
Duca degli Abruzzi 24, 10129 Torino, Italy}

\abst{
We consider the relation between the Boltzmann temperature
and the Lagrange multipliers associated with energy average
in the nonextensive thermostatistics. In Tsallis' canonical
ensemble, the Boltzmann temperature depends on energy
through the probability distribution unless $q=1$. It is
shown that the so-called 'physical temperature' introduced
in [Phys. Lett. A \textbf{281} (2001) 126] is nothing but
the ensemble average of the Boltzmann temperature. }

\begin{document}

\maketitle

\section{Introduction}
It is well-known, particularly in thermal physics
\cite{Callen}, that temperature is a fundamental concept,
and there are some different definitions based on: e.g.,
the zeroth
law of thermodynamics (equilibrium temperature); the Gibbs
distribution (the inverse of the Lagrange multiplier
associated with the energy average); the relation 
between heat and Clausius' entropy;
the Boltzmann temperature
and so on. The important fact is that, in the standard
thermostatistics, all these temperatures become
macroscopically equivalent one another, at equilibrium.\\
In the last years there have been some different approaches to define
temperature in the nonextensive thermostatistics framework
\cite{Tsallis88, Tsallis98, book} based on the above definitions valid
in the standard thermostatitistics.
To cite a few,  the so-called
'physical temperature' \cite{Abe01,Toral03} defined through
the zeroth law of thermodynamics; the `distribution
temperature' defined through the deformed exponential type
distribution \cite{Naudts@NEXT03} $p(E) = \exp_{\phi} (G -
E/T)$, where $\phi$ stands for a set of deformed parameters
of a generalized exponential function $\exp_{\phi}(x)$ and
$G$ accounting for the probability normalization and, more recently,
the temperature
defined through the relationship between heat and Clausius'
entropy in quasi-static process\cite{Abe05}. However, in the generalized
thermostatistics, which describes the thermal properties of
a nonextensive system with
long-range interaction or long-time correlation 
(memory) in non-equilibrium, temperatures with different definitions
appear to be different each other.\\
The purpose of the present contribution is to 
study the Boltzmann temperature
in the non extensive scenarios, which is never discussed before in the 
literature as far as our best knowledge. 
It is shown that the ensemble average of this
temperature is equivalent to the 'physical 
temperature' introduced in\cite{Abe01,Toral03}.

Let us recall that, in  the standard thermostatistics, the
Boltzmann temperature $T^{\rm B}$ introduced as the energy
derivative of the Boltzmann entropy $S^{\rm B} \equiv \ln
\Omega(E)$, [throughout this paper the Boltzmann constant
is set to unity ($k_{\rm B}=1$)]
\begin{equation}
   \frac{1}{T^{\rm B}} \equiv \frac{d \ln \Omega(E)}{dE},
\end{equation}
with $\Omega(E)$ the number of accessible states between
the energies $E$ and $E+\delta E$ of the system.\\ We
recall that the relevant probability distribution of a
system, as a function of the energy level $E$, can be
written as $P(E)=\Omega(E)\,p(E)$\cite{Callen}, where
$p(E)$ is the statistical factor, i.e., a probability of a
single state with energy $E$, whereas $\Omega(E)$ is the
multiplicity of such states. In the Gibbs case, the
statistical factor is given by $p^{\rm G}(E) = Z(\beta^{\rm
G})^{-1} \exp(- \beta^{\rm G}\,E)$. We observe that the
most probable state of a system is the one which maximizes
$P(E)$, i.e., the product $\Omega(E) p(E)$. Let us denotes
$E^*$ the energy at the maximum, then
\begin{equation}
  \frac{d}{dE} \Omega(E)p(E) \Big\vert_{E=E^*} = 0.
\end{equation}
This leads to
\begin{equation}
  -\frac{d \ln p(E)}{dE}\Big\vert_{E=E^*}
  = \frac{d \ln \Omega(E)}{dE} \Big\vert_{E=E^*}
    \equiv\frac{1}{T^{\rm B}(E^*)}.\label{TB}
\end{equation}
For the Gibbs canonical distribution $p^{\rm G}(E) \propto
\exp(-\beta^{\rm G} E)$, the above relation implies the
well-known relation $\beta^{\rm G}=1/T^{\rm B}(E^*)$. 
However, the Lagrange
multiplier for a generalized probability distribution
$p(E)$, is not necessarily equal to the inverse Boltzmann
temperature!

The plane of the paper is as follow. In the next section we
evaluate the Boltzmann temperature at the most probable
state within the $S_{2-q}$ formalism, which is developed in
our previous paper \cite{Wada-Scarfone05}. Section 3
discusses the $q$-Gaussian \cite{Suyari04} approximation of a positive
function with one maximum. In section 4, we apply
this approximation to evaluate the ensemble average of the
inverse Boltzmann temperature. The final section is our
conclusion.

\section{The Boltzmann temperature in the $S_{2-q}$ formalism}
Let us begin with the statistical factor of Tsallis type
\cite{Wada-Scarfone05,Wada-Scarfone}
\begin{equation}
  p(E) = \alpha\qexp \left(-\beta E -\gamma \right),
 \label{pdf}
\end{equation}
where
\begin{equation}
  \qexp(x) \equiv \big[1+(1-q) x \big]^{\frac{1}{1-q}},
\end{equation}
is the Tsallis $q$-exponential and its inverse function,
the $q$-logarithmic, is defined by
\begin{equation}
  \qln(x) \equiv \frac{x^{1-q}-1}{1-q}.
\end{equation}
In Eq. \eqref{pdf} $\beta$ is the Lagrange multiplier
associated with the energy expectation value
\begin{equation}
  U= \int_0^{\infty} dE \; E \Omega(E) p(E),
\end{equation}
and $\gamma$ is the one associated with the probability
normalization
\begin{equation}
  1 = \int_0^{\infty} dE \Omega(E) p(E),
\end{equation}
The parameter $\alpha$ is choosed through the relation
\begin{equation}
   \frac{d}{dx} \Big( x \qln x \Big) = \qln \left( \frac{x}{\alpha} \right),
   \label{prop}
\end{equation}
so that
\begin{equation}
   \frac{1}{\alpha} = (2-q)^{\frac{1}{1-q}}.
   \label{alpha}
\end{equation}
From Eq. \eqref{pdf}, we have
\begin{equation}
  \qln \left( \frac{p(E)}{\alpha} \right)
  = -\beta E-\gamma.
  \label{scaled-log}
\end{equation}
This relation and the property \eqref{prop} guarantee that
Eq. \eqref{pdf} is the solution of the following MaxEnt
problem \cite{Wada-Scarfone05}
\begin{equation}
 \frac{\delta}{\delta p(E)}
 \left(
    S_{2-q} - \beta \int_0^\infty dE \Omega(E) p(E) E -
    \gamma \int_0^\infty dE \Omega(E) p(E)
 \right) = 0,
  \label{MaxEnt}
\end{equation}
where
\begin{equation}
 S_{2-q}= \int_0^\infty dE \Omega(E)  \frac{ \big[p(E)\big]^{2-q} -p(E)}{q-1} \ =
-\int_0^\infty dE \Omega(E) p(E) \qln \left( p(E) \right),
  \label{S_2-q}
\end{equation}
is the Tsallis entropy with $q$ replaced by $2-q$.\\
From Eq. \eqref{scaled-log} we see that the Lagrange
multiplier $\beta$ can be determined by
\begin{equation}
  -\frac{d}{d E} \qln \left( \frac{p(E)}{\alpha} \right)
    = \beta.
\end{equation}
Note that this relation holds for all energies, i.e.
$\beta$  does not dependent on the energy spectrum.\\
By utilizing the identity $\qln (x y) = \qln(x) + \qln(y)
\big( 1+(1-q)\qln(x) \big)$, and Eq. \eqref{alpha}, it
follows
\begin{equation}
  \qln \left( \frac{p(E)}{\alpha} \right) = 1 + (2-q) \qln p(E).
\end{equation}
We have
\begin{align}
  -\beta= (2-q) \frac{d \qln p(E)}{dE}= (2-q) \big[p(E)\big]^{1-q} \frac{d \ln p(E)}{dE} .
\end{align}
Consequently, according to Eq. (\ref{TB}), it follows
\begin{equation}
  \frac{1}{T^{\rm B}(E^*)} = -\frac{d \ln p(E)}{dE} \Big\vert_{E=E^*}
 = \frac{\beta}{2-q} \big[ p(E^*) \big]^{q-1},
  \label{inv TB}
\end{equation}
which relates the Boltzmann temperature $T^{\rm B}(E^*)$,
evaluated at the most probable energy level $E^\ast$, to the
Lagrange multiplier $\beta$. In the $q \to 1$ limit, it
reduces to the standard case $1/T^{\rm B} = \beta$.

Now, let us evaluate $T^{\rm B}(E^*)$. We observe that, by
taking the average of the both sides of Eq.
\eqref{scaled-log} we obtain the following relation
\cite{Wada-Scarfone05}
\begin{equation}
  \gamma+1 = (2-q) S_{2-q}- \beta U.\label{gamma}
\end{equation}
Utilizing the fact that $E^*$ is almost equal to the
average energy $U$ (see Appendix A), we have
\begin{align}
  \big[ p(E^*) \big]^{1-q} &= \alpha^{1-q}\big[1-(1-q)(\beta E^\ast + \gamma)\big]\nonumber\\
  &=1- \left(\frac{1-q}{2-q}\right)
         \Big[ \beta (E^*-U) + (2-q) S_{2-q} \Big] \nonumber \\
  &\approx 1+(q-1) S_{2-q} = \int_0^\infty dE \Omega(E)
  \big[p(E)\big]^{2-q},\label{tt}
\end{align}
where we employed Eq. (\ref{gamma}). From the
definition (\ref{inv TB}) we then obtain
\begin{equation}
  T^{\rm B}(E^*) \approx
   \frac{2-q}{\beta} \int_0^\infty dE
     \Omega(E) \big[ p(E) \big]^{2-q}.\label{TB1}
\end{equation}
Finally, recalling that\cite{Wada-Scarfone05} $d S_{2-q}/dU
= \beta$, Eq. (\ref{TB1}) can be rewritten in the form
\begin{equation}
  \frac{T^{\rm B}(E^*)}{2-q} \approx
   \Big[ 1+(q-1) S_{2-q} \Big]
\left( \frac{d S_{2-q}}{dU} \right)^{-1}.\label{TB2}
\end{equation}
We observe that the r.h.s of Eq. (\ref{TB2}) is equal to
the physical temperature $T_{\rm phys}$, introduced by Abe
\cite{Abe01}, with $q$ replaced by $2-q$.

\section{Approximation of a positive function which has only one maximum}
We consider a positive function $F(x)$ which has only one
maximum point at $x=x^*$ and rapidly decreases as $x$
deviates from $x^*$. In order to approximate $F(x)$ around
$x^*$ let us expand $\qln F(x)$ around $x^*$
\begin{align}
   \qln F(x)
   = \qln F(x^*) &+ (x-x^*) \; \frac{d \qln F(x)}{d x} \Big\vert_{x=x^*}
\nonumber \\
   &+ \frac{1}{2} (x-x^*)^2 \; \frac{d^2 \qln F(x)}{d x^{2}} \Big\vert_{x=x^*}
          + \cdots.
  \label{expanded F}
\end{align}
We choose $x^*$ such that
\begin{equation}
  \frac{d \qln F(x)}{d x}\Big|_{x=x^\ast}
     = \big[ F(x^*) \big]^{-q} \; \frac{d F(x)}{ dx}\Big\vert_{x=x^*} = 0.
  \label{F'}
\end{equation}
Moreover, since $F(x^*)$ is the maximum of $F(x)$, it
should be a concave function $d^2 F(x) /
dx^2\big|_{x=x^\ast} < 0$. Introducing the positive
quantity $\sigma_{x^*}^2$ defined through the relation
\begin{equation}
  \frac{1}{\sigma_{x^*}^2}=
   -\big[F(x^*)\big]^{-1} \; \frac{d^2 F(x)}{ d x^{2}}\Big\vert_{x=x^*},
\end{equation}
we have
\begin{equation}
  \frac{d^2 \qln F(x)}{d x^{2}} \Big\vert_{x=x^*}
     = \big[ F(x^*) \big]^{-q} \; \frac{d^2 F(x)}{ dx^2} \Big\vert_{x=x^*}
     = -\frac{\big[ F(x^*) \big]^{1-q}}{\sigma_{x^*}^2}.
  \label{F''}
\end{equation}
Substituting Eqs. \eqref{F'} and \eqref{F''} into Eq.
\eqref{expanded F}, and utilizing the identity
\begin{equation}
  \qln \left(\frac{f}{g} \right) = \frac{\qln f - \qln g}{g^{1-q}},
\end{equation}
it follows
\begin{equation}
 \qln \left( \frac{F(x)}{F(x^*)} \right) \approx
       -\frac{(x-x^*)^2}{2 \sigma_{x^*}^2}.
\end{equation}
Thus $F(x)$ can be well approximated by the $q$-Gaussian
function \cite{Suyari04}
\begin{equation}
 F(x) \approx F(x^*)\; \qexp\left( -\frac{(x-x^*)^2}{2 \sigma_{x^*}^2}
 \right).\label{appr}
\end{equation}
Remark that such approximation is more and more better when
$\sigma_{x^*}^2 \ll 1$.

Now let us apply the above $q$-Gaussian approximation to
the non self-referential expression of the $p(E)$ given by
Eq. \eqref{pdf}. Putting $F(E) \equiv P(E) = \Omega(E)
p(E)$, then
\begin{equation}
 \qln P(E) = \qln \Omega(E) -\frac{1}{2-q} \left( \beta E+\gamma+1\right)
      \big[\Omega(E)\big]^{1-q}.
  \label{qln F}
\end{equation}
Remark that $E^*$ satisfies
\begin{equation}
  \frac{d \qln P(E)}{d E} \Big\vert_{E=E^*}= 0,
\end{equation}
which leads to Eq. \eqref{inv TB} as shown in Appendix B.
The second derivative becomes (see also Appendix B)
\begin{align}
  -\frac{1}{\sigma_{E^*}^2} = \frac{d^2 \ln \Omega(E)}{d E^{2}} \Big\vert_{E=E^*}
     -(1-q) \left( \frac{d \ln \Omega(E)}{d E} \Big\vert_{E=E^*}\right)^2.
  \label{2nd derivative}
\end{align}
In the standard case ($q=1$) Eq. \eqref{2nd derivative}
reduces to the concavity condition for the Boltzmann
entropy $d^2 \ln \Omega(E) / d E^{2} \big\vert_{E=E^*} <
0$. However, in nonextensive case ($q \ne 1$), the r.h.s of
Eq. (\ref{2nd derivative}) becomes negative even for a
convex function $\ln\Omega$, if $q<1$ and
\begin{equation}
 0< \frac{d^2 \ln \Omega(E)}{d E^{2}} \Big\vert_{E=E^*} <
     (1-q) \left( \frac{d \ln \Omega(E)}{d E} \Big\vert_{E=E^*} \right)^2.
\end{equation}
Since
\begin{equation}
\frac{d^2 \ln_{2-q} \Omega(E)}{d E^{2} }
   = \left[ \frac{d^2 \ln \Omega(E)}{d E^{2}}
     -(1-q) \left( \frac{d \ln \Omega(E)}{d E} \right)^2 \right] \big[\Omega(E)\big]^{q-1},
\end{equation}
Eq. \eqref{2nd derivative} can be rewritten as
\begin{equation}
  -\frac{1}{\sigma_{E^*}^2} = \left[\Omega(E^*)\right]^{1-q}
  \frac{d^2 \ln_{2-q} \Omega(E)}{d E^{2} } \Big\vert_{E=E^*}.
\end{equation}
Introducing the $q$-generalized Boltzmann entropy
\begin{equation}
  S_{2-q}^{\rm B}(E) \equiv \ln_{2-q} \Omega(E),
\end{equation}
the condition of the positivity of $\sigma_{E^*}^2$ becomes
\begin{equation}
  \frac{d^2 S_{2-q}^{\rm B}(E)}{d E^{2} } \Big\vert_{E=E^*} < 0,
\end{equation}
which is nothing but the concavity condition of
$S_{2-q}^{\rm B}(E)$.

\section{The ensemble average of the inverse Boltzmann temperature}
Let us evaluate the ensemble average of the inverse
Boltzmann temperature. We assume that $P(E)=\Omega(E) p(E)$
is well approximated by $q$-Gaussian function as in Eq.
\eqref{appr}. Taking logarithm of the both sides of Eq.
\eqref{appr} we obtain
\begin{equation}
\ln \Omega(E) \approx \ln \Omega(E^*) + \ln p(E^*) - \ln
p(E)
      +\ln \left( \qexp\left( -\frac{(E-E^*)^2}{2 \sigma_{E^*}^2} \right)
  \right).
\end{equation}
The ensemble average of the inverse Boltzmann temperature
becomes
\begin{align}
 \ave{\frac{1}{T^{\rm B}(E)}} &= \ave{ \frac{\ln \Omega(E)}{dE}}
\nonumber \\
 &\approx \ave{-\frac{\ln p(E)}{dE}}
-\ave{\frac{(E-E^*)}{ \sigma_{E^*}^2}
 \left[ \qexp\left( -\frac{(E-E^*)^2}{2 \sigma_{E^*}^2} \right)
 \right]^{q-1}}.\label{TB3}
\end{align}
The second term in the r.h.s. is proportional to
\begin{align}
   \int_0^\infty dE (E-E^*)
  \left[ \qexp\left( -\frac{(E-E^*)^2}{2 \sigma_{E^*}^2} \right)
  \right]^q,
\end{align}
and changing the integration variable $E$ to $u=E-E^*$ it
becomes
\begin{align}
 \int_{-E^*}^{\infty} du \; u
 \left[ \qexp\left( -\frac{u^2}{2 \sigma_{E^*}^2} \right) \right]^q.
  \label{int}
\end{align}
As similar as Appendix A we assume that $E^* \gg 1$ and
$\sigma_{E^*}^2\ll1$. Under these conditions, the integral
\eqref{int} is well approximated with
\begin{align}
 \int_{-\infty}^{\infty} du \; u
 \left[ \qexp\left( -\frac{u^2}{2 \sigma_{E^*}^2} \right)
 \right]^q,
\end{align}
which vanishes. Consequently, from Eq. (\ref{TB3}), it
follows
\begin{align}
 \ave{\frac{1}{T^{\rm B}(E)}}
 \approx \frac{\beta}{2-q} \ave{\big[ p(E) \big]^{q-1}}
 = \frac{\beta}{2-q} \int_0^{\infty} dE \Omega(E) \big[p(E)\big]^q.
\end{align}
By utilizing the following relation \cite{Wada-Scarfone05}
 between $\beta$ and $\beta^{\rm (3)}$
\begin{align}
 \beta^{\rm (3)}
 = \frac{\beta}{2-q}
   \left( \int_0^{\infty} dE \Omega(E) \big[p(E)\big]^q \right)^2,
\end{align}
where $\beta^{(3)}$ is the Lagrange multiplier associated with the
energy expectation value in the third formalism \cite{Tsallis98}, 
we finally obtain
\begin{align}
 \ave{\frac{1}{T^{\rm B}(E)}}
 \approx \frac{\beta^{(3)}}{\int_0^\infty dE\Omega(E)\big[p(E)\big]^q}=
 \frac{\beta^{\rm (3)}}{1+(1-q)S_q} = \frac{1}{T_{\rm phys}}.
\end{align}
Thus the physical temperature \cite{Abe01}, which follows based on 
the thermodynamics zeroth law argument, is almost equal
to the ensemble average of the Boltzmann temperature.

\section{Conclusion}
We have considered the relationship between the Boltzmann
temperature and the Lagrange multipliers associated with
the energy average in the nonextensive thermostatistics. It
is shown that the so-called 'physical temperature' is
nothing but the ensemble average of the Boltzmann
temperature. In the Tsallis canonical ensemble, unless
$q=1$, the Boltzmann temperature depends on energy through
the probability distribution, i.e., it fluctuates with
energy fluctuations.


\appendix
\section{The derivation of $E^* \approx U$}
According to Eq. (\ref{appr}) we assume that $F(E)\equiv
P(E)=\Omega(E) p(E)$ is well approximated by the
$q$-Gaussian function
\begin{equation}
 \Omega(E)p(E) \approx \Omega(E^*)p(E^*)
      \; \qexp\left( -\frac{(E-E^*)^2}{2 \sigma_{E^*}^2} \right).
  \label{q-Gauss}
\end{equation}
Let us consider the average of $E-E^*$
\begin{align}
 U - E^* &= \ave{E - E^*} = \int_0^{\infty} dE (E-E^*) \Omega(E) p(E)
\nonumber \\
 &\approx \Omega(E^*)p(E^*)
      \int_0^{\infty} dE (E-E^*)
 \qexp\left( -\frac{(E-E^*)^2}{2 \sigma_{E^*}^2}
 \right).\label{b2}
\end{align}
By changing the integration variable $E$ to $u=E-E^*$, the
integral (\ref{b2}) becomes
\begin{align}
U-E^\ast \int_{-E^*}^{\infty} du \; u
 \qexp\left( -\frac{u^2}{2 \sigma_{E^*}^2} \right).
\end{align}
Observing that $E^*$ is a macroscopic quantity, i.e., $E^*
>> 1$ and because $\sigma_{E^*}^2$ should be a small quantity, the integral
can be well approximated with
\begin{align}
U-E^\ast= \int_{-\infty}^{\infty} du \; u
 \qexp\left( -\frac{u^2}{2 \sigma_{E^*}^2} \right),
\end{align}
which vanishes. As a consequently $E^* \approx U$.

\section{The derivation of Eq. \eqref{2nd derivative}}

From Eq. \eqref{qln F}, it follows
\begin{align}
 0 &= \frac{1}{\Omega(E^*)^{1-q}} \;
       \frac{d \qln P(E)}{d E} \Big\vert_{E=E^*} \nonumber \\
    &= \frac{d \ln \Omega(E)}{d E}  \Big\vert_{E=E^*}
          \left[ 1-\left( \frac{1-q}{2-q} \right)
                \left( \beta E^*+\gamma+1\right)
          \right] - \frac{\beta}{2-q},\label{b1}
\end{align}
which is equivalent to Eq. \eqref{inv TB}. Next, let us
evaluate the second derivative. We see
\begin{align}
 \frac{1}{\Omega(E^*)^{1-q}} \; \frac{d^2 \qln P(E)}{d E^{2}} \Big\vert_{E=E^*}
    &= \frac{d^2 \ln \Omega(E)}{d E^2} \Big\vert_{E=E^*}
          \left[ 1-\left( \frac{1-q}{2-q} \right)
                \left( \beta E^*+\gamma+1 \right)
          \right] \nonumber \\
     &\qquad - (1-q) \frac{d \ln \Omega(E)}{d E} \Big\vert_{E=E^*}
           \left( \frac{\beta}{2-q} \right),
  \label{F2d}
\end{align}
and
\begin{align}
 \left[{P(E^*)\over\Omega(E^\ast)}\right]^{1-q} =
     1-\left( \frac{1-q}{2-q} \right)
                \left( \beta E^*+\gamma+1\right).
  \label{F^1-q}
\end{align}
Dividing Eq. \eqref{F2d} by Eq. \eqref{F^1-q}, and accounting
for Eq. (\ref{b1}), it leads to Eq. \eqref{2nd derivative}.

\end{document}